\documentclass{sig-alternate-05-2015}

\usepackage{xcolor}
\usepackage{breakurl}

\usepackage[noend]{algorithmic}
\usepackage{algorithm}
\usepackage[nospace,sort,compress]{cite}
\usepackage{microtype}
\usepackage{balance}

\newcommand{\VTx}{\mbox{VT-x}}
\newcommand{\VTd}{\mbox{VT-d}}

\begin{document}

\toappear{
\copyright{} held by the authors (2017). 
This is the author's extended version of the work. 
It is intended for your personal use. 
Not for redistribution. 
The definitive version will be published at\\
\textbf{CODASPY'17}, {March 22 -- 24, 2017, Scottsdale, AZ, USA} \\
ISBN: 978-1-4503-4523-1/17/03\\
DOI: \url{http://dx.doi.org/10.1145/3029806.3029822}
}

\clubpenalty=10000
\widowpenalty = 10000

\title{SGXIO: Generic Trusted I/O Path for Intel SGX}

\numberofauthors{2}
\author{
\alignauthor
Samuel Weiser\\
       \affaddr{Graz University of Technology, Austria}\\
       \email{samuel.weiser@iaik.tugraz.at}
\alignauthor
Mario Werner\\
       \affaddr{Graz University of Technology, Austria}\\
       \email{mario.werner@iaik.tugraz.at}
}

\maketitle

\begin{abstract}

Application security traditionally strongly relies upon security of the underlying operating system. 
However, operating systems often fall victim to software attacks, compromising security of applications as well. 
To overcome this dependency, Intel introduced SGX, which allows to protect application code against a subverted or malicious OS by running it in a hardware-protected enclave. 
However, SGX lacks support for generic trusted I/O paths to protect user input and output between enclaves and I/O devices. 

This work presents SGXIO, a generic trusted path architecture for SGX, allowing user applications to run securely on top of an untrusted OS, while at the same time supporting trusted paths to generic I/O devices. 
To achieve this, SGXIO combines the benefits of SGX's easy programming model with traditional hypervisor-based trusted path architectures. 
Moreover, SGXIO can tweak insecure debug enclaves to behave like secure production enclaves. 
SGXIO surpasses traditional use cases in cloud computing and makes SGX technology usable for protecting user-centric, local applications against kernel-level keyloggers and likewise. 
It is compatible to unmodified operating systems and works on a modern commodity notebook out of the box. 
Hence, SGXIO is particularly promising for the broad x86 community to which SGX is readily available.

\end{abstract}


\newcommand{\eg}{{\it e.g.\ }}
\newcommand{\ie}{{\it i.e.\ }}
\newcommand{\cf}{{\it cf.\ }}
\newcommand{\etal}{{\it et al.\ }}
\newcommand{\etalp}{{\it et al. }}
\newcommand{\etalpx}{{\it et al.}}
\newcommand{\etc}{{\it etc.\ }}
\newcommand{\etcp}{{\it etc. }}
\newcommand{\etcpx}{{\it etc.}}
\newcommand{\R}{\textsuperscript{\textregistered}}

\section{Introduction}\label{sec:intro}

Software vulnerabilities are still a predominant issue for application security. 
Over 2000 software vulnerabilities were newly reported within just three months\footnote{This refers to the period between 02/05/2016 and 12/08/2016.}~\cite{nvdsearch}. 
As distributed, networked computing becomes omnipresent in the Internet of Things~(IoT), the risk of damage even increases, allowing remote exploits on a large scale. 

A major reason for this fact is code complexity, which makes traditional secure design paradigms like software verification or testing reach its limits. 
Hence, research focuses on securing sensitive code only and executing it in architecturally isolated containers, often referred to as enclaves. 
Since an enclave is protected against all non-enclave code, the whole OS stack can safely be considered untrusted without reducing security of the enclave. 

The evolution of isolated execution is long and diverse~\cite{overshadow,sp3,ta-min_splitting_2006,mccune_trustvisor:_2010,hofmann_inktag:_2013,
criswell_virtual_2014,suh_aegis:_2003,champagne_scalable_2010,secureme,thekkath_architectural_2000,owusu_oasis:_2013,secureblue,sgxmanual,evtyushkin_iso-x:_2014,costan_sanctum:_2016,trustzone}. 
Intel SGX~\cite{sgxmanual} deserves particular attention since it not only provides comprehensive enclave protection but is also available to the broad Intel x86 community. 
SGX targets high-performance cloud computing, where the cloud provider is entirely distrusted, as well as Digital Rights Management~(DRM).

In order to protect not only enclave execution but also user I/O, one requires trusted paths between enclaves and I/O devices. Currently, SGX only works with proprietary trusted paths like Intel Protected Audio Video Path~(PAVP) which rely on the Intel Management Engine~(ME)~\cite{sgxsoftwaresolutions,ruan_platform_2014}. 
However, proprietary trusted paths are hard to analyze regarding security. 
Moreover, they are not generic and address specific devices and scenarios only. 
Unfortunately, SGX lacks support for generic trusted paths.

\textbf{Contributions.}
In this work we present SGXIO, which is, to our knowledge, the first generic trusted path architecture for SGX.
SGXIO protects secure user applications as well as associated trusted paths against an untrusted OS. 
User applications benefit from SGX protection while trusted paths are established via a small and trusted hypervisor. 
To that end, we identify and solve several challenges in linking the security domains of SGX and the trusted hypervisor. 
This allows a remote party to attest not only enclave code but also the whole trusted path setup. 
Also, SGXIO allows human end users to verify trusted paths without requiring additional hardware.
SGXIO improves upon existing generic trusted paths for x86 systems with an easier and more intuitive programming model.
Furthermore, we show how SGXIO can tweak debug enclaves to behave like production enclaves. 
Therefore, the trusted hypervisor selectively disables enclave debug instructions. 
Finally, we give a novel zero-overhead, non-interactive key transport scheme for establishing an 128-bit symmetric key between two local SGX enclaves. 

The rest of this paper is structured as follows: Section~\ref{sec:related} gives related work, followed by an overview of SGX in Section~\ref{sec:sgx}. 
Section~\ref{sec:problem} discusses the threat model and challenges.
Section~\ref{sec:arch} presents our SGXIO architecture while Section~\ref{sec:security} gives a thorough security analysis. 
Section~\ref{sec:sgx_debug} shows how SGXIO can apply the debug enclave tweak. 
It is followed by further considerations in Section~\ref{sec:implementation} and a conclusion in Section~\ref{sec:conclusion}.


\section{Related Work}\label{sec:related}
This section discusses prior work on isolated execution and trusted paths. Specifically, we compare to ARM TrustZone, which is ARM's counterpart to SGX.

\subsection{Isolated Execution}

Traditionally, security kernels are used to achieve strong process and resource isolation. 
One soon realized that security kernels itself are the Achilles' heel of the whole system security and need complexity reduction. 
This paved the way for microkernels, for which seL4 is a prominent example. 
The developers of seL4 reduced kernel code size to 8700 lines of C code and conducted formal verification to proof correctness~\cite{sel4proof}. 
Even with a secure microkernel in place, designing a fully-featured, secure OS on top of it is still an unmet challenge~\cite{jacobsen_lightweight_2016}. 
Hence, one has to live with a feature-rich but untrusted OS. 

Isolating an application from such an untrusted OS is essential for any trusted path. 
There exist various techniques for isolated execution, which range from pure hypervisor designs~\cite{overshadow,sp3,ta-min_splitting_2006,mccune_trustvisor:_2010,hofmann_inktag:_2013,criswell_virtual_2014} over hardware-software co-designs~\cite{suh_aegis:_2003,champagne_scalable_2010,secureme} to pure hardware extensions~\cite{thekkath_architectural_2000,owusu_oasis:_2013,secureblue,sgxmanual,evtyushkin_iso-x:_2014,costan_sanctum:_2016,trustzone}. 
SGXIO uses Intel SGX~\cite{sgxmanual}, which, from a functional perspective, cumulates previous work in software isolation, attestation and transparent memory encryption. 
Others isolate sensitive code by keeping it entirely in the CPU cache~\cite{carma} or by migrating it to system management RAM on x86~\cite{azab_sice:_2011}. Often,  dedicated security co-processors are used~\cite{smith_building_1999}. 

The Trusted Platform Module~(TPM)~\cite{tpm} is a security co-processor which does not directly offer software isolation on its own. However, it can be used in conjunction with Intel TXT~\cite{inteltxtmanual} to set up an isolated execution environment~\cite{mccune_flicker:_2008}.

Any communication with the untrusted OS is problematic and needs careful validation~\cite{ports_towards_2008,checkoway_iago_2013}. 
Baumann \etal therefore reduce the untrusted syscall interface to a bare minimum and shift a whole Windows 8 library into an SGX enclave, which tremendously increases the TCB~\cite{haven}. 

\subsection{Trusted Paths}

There exist various attempts for integrating trusted paths directly into existing commodity OSes. However, they usually suffer from a bloated TCB, covering the whole OS~\cite{liu_screenpass:_2013,tong_guardroid:_2013,lange_crossover:_2013,fernandes_tivos:_2014}.
More sound trusted paths consider the OS untrusted. One can distinguish between generic trusted paths and specific trusted paths. Latter are limited to specific devices or scenarios. 

\textbf{Generic Trusted Paths.}
Zhou \etal build a generic trusted path on x86 systems in a pure hypervisor-based design~\cite{zhou_building_2012}. 
They show the first comprehensive approach on x86 systems, protecting a trusted path all the way from the application level down to the device level. 
They consider PCI device misconfiguration, DMA attacks as well as interrupt spoofing attacks. 
However, pure hypervisor-based designs come at a price. 
They strictly separate the untrusted stack from the trusted one. 
Hence, the hypervisor is in charge of managing all secure applications and all associated resources itself. 
This includes secure process and memory management with scheduling, verified launch and attestation. 
Also, communication between both security domains might be non-trivial due to synchronization issues or potentially mismatching Application Binary Interfaces~(ABI).
In contrast, SGXIO uses the comparably easy programming model of SGX enclaves, in which the untrusted OS is in charge of managing secure enclaves. Moreover, SGX provides verified launch and attestation out of the box.

\textbf{Specific Trusted Paths.}
In~\cite{zhou_-demand_2014} and~\cite{zhou_dancing_2014}, Zhou \etal discuss  a trusted path to a single USB device. Yu \etal show how to apply the trusted path concept to GPU separation~\cite{yu_trusted_2015}.
Filyanov \etal discuss a pure uni-directional trusted path using the TPM and Intel TXT~\cite{filyanov_uni-directional_2011}, which has two notable drawbacks: 
First, it is limited to trusted input paths only. 
Second, while waiting for secure input from the user, the OS is suspended. 
In contrast, SGXIO supports full parallelism between the untrusted OS and secure applications.

Another way of establishing a trusted path is via dedicated I/O devices~\cite{perrig_safe_2009,pci_security_standards_council_approved_????}. 
The I/O device natively supports cryptography, which allows an application to directly open a cryptographic channel to it. 
This bypasses any untrusted software or hardware, making a trusted hypervisor unnecessary. 
However, this concept does not generalize to other I/O devices, especially legacy devices. 
 
Many trusted paths build on proprietary hardware and software like Intel's Protected Audio Video Path~(PAVP) as well as its successor, Intel Insider~\cite{ruan_platform_2014,knupffer_intel_2011}. 
Both rely on Intel's proprietary Management Engine~(ME).
Hoekstra \etal outline integration of PAVP in SGX applications to achieve secure video conferencing and one-time password generation~\cite{sgxsoftwaresolutions}. 
However, they do not come up with any trusted input path solution, deferring this as future work. 
In~\cite{ruan_platform_2014}, Ruan describes a Protected Transaction Display~(PTD) application running on the ME, which makes use of PAVP to securely obtain a one-time PINs from the user. 
However, ME-related code is proprietary and kept under wraps. 
This not only strongly limits potential use cases but also hinders transparent security assessment.
 
\subsection{ARM TrustZone}
ARM TrustZone combines secure execution with trusted path support. 
A TrustZone compatible CPU provides a secure world mode, which is orthogonal to classical privilege levels. 
The secure world is isolated against the normal world and operates a whole trusted stack, including security kernel, device drivers and applications. 
In addition, TrustZone is complemented by a set of hardware modules, which allow strong isolation of physical memory as well as peripherals. 
Also, device interrupts can be directly routed into the secure world. 
TrustZone can be combined with a System MMU, similar to an IOMMU, which can prevent DMA attacks. 
Thus, TrustZone not only allows isolated execution~\cite{sun_trustice:_2015} but also generic trusted paths~\cite{li_building_2014}, which is a significant advantage over SGX. 
In contrast to SGX, TrustZone does not distinguish between different secure application processes in hardware. 
It requires a security kernel for secure process isolation, management, attestation and similar.  


\section{Software Guard Extensions}\label{sec:sgx}
Intel Software Guard Extensions~(SGX) have been rolled out with Skylake in October 2015~\cite{sgxshipping}. SGX comprises a new set of x86 instructions, enabling user applications to declare parts of its virtual address space as secure enclave. The enclave can access its hosting application's memory while the host cannot touch enclave memory. In general, any non-enclave access into the enclave is prohibited by the CPU. 
The OS is entirely distrusted and is supervised by the CPU in all enclave management operations. In addition, SGX encrypts all enclave memory on the fly when written to DRAM using a dedicated hardware encryption module. SGX provides verified enclave launching, attestation and sealing. As such, SGX encourages a small Trusted Computing Base~(TCB), only consisting of enclave code and the CPU itself. 

In the following we outline verified enclave launch, attestation and sealing before discussing enclave debugging and licensing. 
For more details we refer to available literature~\cite{sgxmanual,sgxsdm,sgxtutorial,costan_intel_2016,mckeen_innovative_2013,anati_innovative_2013,sgxsoftwaresolutions,sgxsdk,sgxdevguide}. Especially dynamic enclave page management, which is not covered here, can be looked up in~\cite{mckeen_innovative_2013,sgxmanual}.

\textbf{Verified Enclave Launch.}\label{sec:enclave_startup}
When loading an enclave into memory, the CPU measures its content in a chained cryptographic hash log, stored in a register called \verb!MRENCLAVE!. This is comparable to a Platform Configuration Register~(PCR) in a TPM~\cite{tpm}. Before running the enclave, the CPU verifies \verb!MRENCLAVE! against a vendor-signed version and aborts on a mismatch. Hence, \verb!MRENCLAVE! vouches for integrity of the enclave startup.

\textbf{Attestation and Sealing.}
In order to assess enclave security, SGX provides attestation mechanisms~\cite{anati_innovative_2013}. Local attestation enables an enclave to verify another enclave running on the same physical CPU. Remote attestation can be used by a remote party to check if the attested enclave is indeed running on a genuine Intel CPU. It allows initial provisioning of keys and secrets. This is required since enclave code is public, not allowing to embed secrets directly in the code. Attestation is based on a CPU-generated, signed report structure containing \verb!MRENCLAVE!. The report structure is able to hold additional user data. One can use this to authentically exchange information between enclaves via local attestation and agree on an encryption key, for example. 

SGX also allows an enclave to obtain a sealing key which is bound to the local CPU. SGX permits making the sealing key dependent on \verb!MRENCLAVE!. Hence, the same sealing key can only be queried from exactly the same enclave, if loaded correctly on the same, genuine Intel CPU. The enclave can use the sealing key to encrypt arbitrary data for offline storage, preserving its state among multiple system reboots. 

\textbf{Debugging.}
SGX distinguishes between debug and production enclaves. Debug enclaves can be accessed by the OS via \verb!EDBGRD! and \verb!EDBGWR! instructions while production enclaves cannot. Moreover, debug enclaves can opt-in to ordinary x86 breakpoint handling and performance monitoring~\cite{sgxsdm}. This supports the enclave development process. In a production setting, however, enclaves have to run in production mode to protect against an untrusted OS. During initialization of an enclave, one can set a debug mode flag, specifying whether the enclave shall be run in debug or production mode. This choice yields different \verb!MRENCLAVE! values for either option, making a debug enclave distinguishable from a production enclave. 

\textbf{Enclave Licensing.}
SGX has a controversially discussed "feature", called launch enclave~\cite{beekman_intel_2015}. During enclave initialization, the CPU verifies a so-called \verb!EINITTOKEN!, which contains several enclave attributes to enforce, including the debug mode flag. The \verb!EINITTOKEN! has to be signed by a special launch key, which is owned by so-called launch enclaves, issued by Intel. Hence, by issuing proper launch enclaves, Intel has full control over which enclaves are to be executed in debug or production mode. The SGX evaluation SDK is shipped with a launch enclave issuing \verb!EINITTOKEN!s for debug enclaves only~\cite{sgxsdk}. In order to run an enclave in production mode, one needs to obtain a proper license from Intel~\cite{johnson_intel_2016}.


\section{Threat Model and Challenges}\label{sec:problem}

SGXIO utilizes SGX as one building block to provide isolated execution. 
However, the threat models of pure SGX and SGXIO differ. 
This section elaborates on the threat model of SGXIO and shows that, in contrast to pure SGX, physical attacks don't have to be considered for trusted paths. 
Furthermore, the challenges arising from the combination of SGX with a trusted hypervisor are discussed.

\subsection{Distinction from SGX}\label{sec:use_case}

\textbf{SGX} has been designed as isolated execution technology with a minimal trusted computing base~(TCB). 
The TCB only contains the processor itself, which acts as trust anchor, and the code running within enclaves. 
Everything else is considered potentially malicious. 
This not only includes all other software components (e.g.,~OS, hypervisor) but also the hardware environment it operates in. 
Therefore, SGX not only considers logical attacks but also physical attacks.

This threat model perfectly fits the requirements for secure cloud computing in which a customer wants to protect enclave code and data against an untrusted cloud provider, controlling the software stack and the hardware. 
In this use case, all communication with an enclave can be performed using securely encrypted and authenticated channels.
Also, content providers can use SGX to enforce a DRM scheme on an untrusted consumer PC.

In a local setting, however, a user wants to benefit from SGX by protecting user-centric applications against a potentially compromised OS. 
Especially, the communication between user apps and the user via I/O peripherals needs protection from the OS via a trusted path. 
This setting somehow contradicts the threat model of SGX, which considers the physical environment, and therefore also the local user, a threat.  
Currently, in order to achieve a trusted path with SGX, one has to rely on encrypted interfaces like PAVP. 
However, the prevalence of unencrypted I/O devices in todays computers and the lack of support to securely communicate with these devices demands other, more generic mechanisms. 

\begin{figure}
\centering
\includegraphics[width=0.5\textwidth]{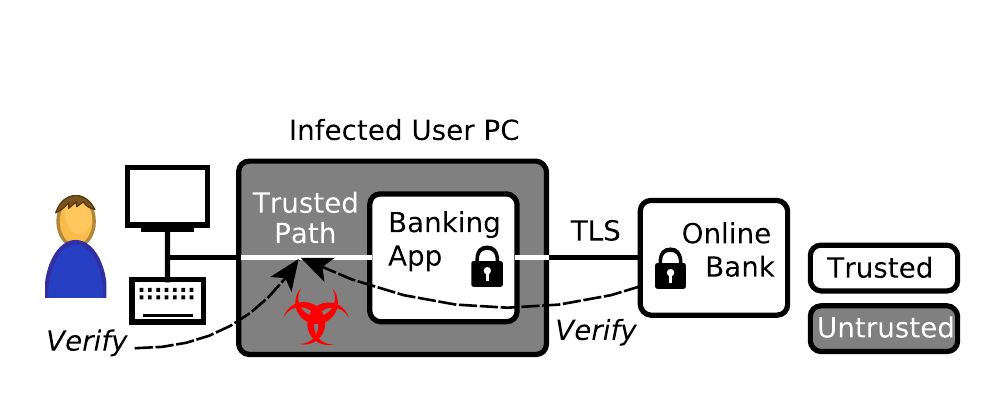}
\caption{In an online banking scenario, malware shall not be able to hijack a banking session. 
Communication with the bank is encrypted via TLS, while the banking app itself is protected with SGX. 
Both the user and the bank want to be able to verify security of the trusted path. }
\label{fig:banking}
\end{figure}

\textbf{SGXIO} fixes this shortcoming by extending SGX with a generic trusted path. 
Many user-centric applications can profit from this additional feature. 
This covers pure local applications like confidential document viewers, anti-spoofing password prompts, secure password generators and password safes but also internet scenarios like secure conferencing and chat applications as well as secure online banking. 
To take latter as example, online banking can not only be secured up to the user's browser via TLS, for example, but up to the I/O devices via trusted paths, as depicted in Figure~\ref{fig:banking}. 
Moreover, SGXIO allows both the user and the online bank to verify the trusted path. 
This means that sensitive information like login credentials, the account balance, or the transaction amount can be protected even if other software running on the user's computer, including the OS, is infected by malware. 
Additionally, SGXIO provides attestation mechanisms to enable the bank as well as the user to verify that trusted paths are established and functional.

Having a trusted path has implications on the threat model of SGXIO. 
A physical attacker has direct access to I/O devices and can impersonate the user without subverting trusted paths. 
Thus, trusted paths can only protect against logical attacks but cannot provide physical security at all. 
The following section explains the threat model of SGXIO in detail.

\subsection{Threat Model}\label{sec:threat_models}
In general, the adversaries attacking SGXIO attempt to subvert a trusted path between a user app and an I/O device. 
Subsequently, they succeed if they are able to break the confidentiality or authenticity of such a trusted path.

\textbf{Logical attacks} are the main concern of SGXIO. 
Attackers are assumed to have full control over the OS and know the whole software configuration including all enclave code. 
This is a realistic scenario, addressing both local and remote software attacks which might even yield kernel privileges to attackers. 
Attackers can therefore directly attack enclave interfaces visible to the OS by running enclaves in a fake environment within the OS. 
Also, attackers can dynamically load and execute custom user apps and drivers and open other trusted path sessions.

Moreover, indirect attacks on a trusted path can be performed by misconfiguring devices under OS control, as outlined by Zhou \etal~\cite{zhou_building_2012}. 
The idea of such attacks is to manipulate noninvolved devices to interfere with a trusted path. 
For example, a PCI devices could be configured such that its address range overlaps those of the user device. 
Also, malicious Direct Memory Access~(DMA) requests could be issued and interrupts could be spoofed.

All code in the trusted computing base (e.g., secure user applications, secure I/O drivers and the hypervisor) is assumed to be correct and not vulnerable to logical attacks. 
Using a formally verified hypervisor such as seL4~\cite{sel4proof} supports this assumption.

\textbf{Physical attacks} are not considered in SGXIO, as already explained, since the user interacting with the system has to be trusted anyway.
As with SGX, Denial-of-Service~(DoS) as well as side channel attacks are also out of scope for SGXIO.

Note that SGXIO requires a modern Intel platform with SGX support as well as support for TPM-based trusted boot. 
All hardware (CPU, chipset, peripherals) is expected to work correctly.

\subsection{Challenges}
SGXIO combines SGX with a trusted hypervisor to provide a generic trusted path. 
However, the hypervisor and SGX form two disjoint security domains with two different trust anchors, which are not designed to collaborate. 
Subsequently, connecting both domains is a non-trivial task. 

This essentially breaks down to two major challenges which had to be solved: 
First, the security domains of the hypervisor and SGX enclaves have to be linked. 
More concretely, we need a way for SGX enclaves to check the presence and the authenticity of the hypervisor. 
We name this problem \emph{hypervisor attestation}. 
Once the hypervisor is attested, it extends trust to any trusted path it establishes. 

Second, the SGXIO architecture relies on multiple SGX enclaves which communicate using keys based on local attestation, as discussed in the following sections. 
These enclaves are executed in different security contexts (trusted hypervisor vs. untrusted OS). 
However, in SGX enclaves are unaware of their context, making them vulnerable to \emph{enclave virtualization attacks}. 
SGXIO prevents such attacks via a careful interface design between both contexts.


\section{SGXIO Architecture}\label{sec:arch}

This section presents our SGXIO architecture and elaborate on its isolation guarantees. 
We discuss design of secure user applications, secure I/O drivers as well as the hypervisor. 

\subsection{Architecture}
\begin{figure}
\centering
\includegraphics[width=0.5\textwidth]{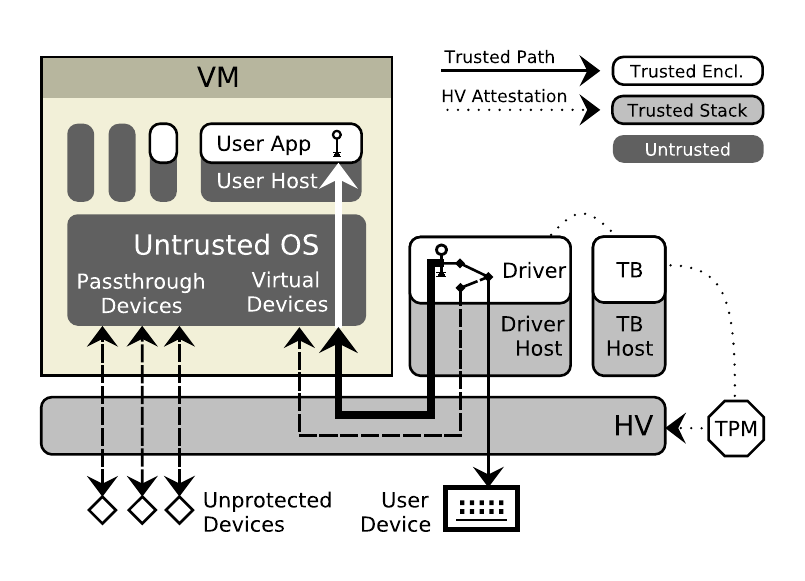}
\caption{The trusted stack consists of a hypervisor~(HV), a Trusted Boot~(TB) enclave and one or more secure I/O drivers. 
The Virtual Machine~(VM) operates an untrusted OS on which secure user apps are hosted. 
The driver obtains data from the user device (thin line) and encrypts it (bold line) for a user app, providing a trusted path (solid line). 
The TB enclave allows drivers to attest the hypervisor.}
\label{fig:sw_stack}
\end{figure}

SGXIO is composed of two parts: a trusted stack and a Virtual Machine~(VM), as seen in Figure~\ref{fig:sw_stack}. 
The trusted stack contains a small security hypervisor, one or more secure I/O drivers, which we simply call \emph{drivers}, as well as a Trusted Boot~(TB) enclave. 
The VM hosts an untrusted commodity OS like Linux, which runs secure user applications, also abbreviated with \emph{user apps}. 

User apps want to communicate securely with the end user. 
They open an encrypted communication channel to a secure I/O driver to tunnel through the untrusted OS. 
The driver in turn requires secure communication with a generic user I/O device, which we term \emph{user device}. 
To achieve this, the hypervisor exclusively binds user devices to the corresponding drivers. 
Note that any other device is directly assigned to the VM. 
I/O on those unprotected devices directly passes through the hypervisor without performance penalty. 
The trusted path names both, the encrypted user-app-to-driver communication and the exclusive driver-to-device binding. 
It is indicated with a solid line in Figure~\ref{fig:sw_stack}. 
Drivers use the TB enclave to get assured of correct trusted path setup by attesting the hypervisor, which is indicated by a dotted line.

\subsection{Isolation Guarantees}\label{sec:isolation}

SGXIO establishes a trusted path all the way from a user app to the user device. 
This requires isolation on several layers. 
First, all trusted stack memory needs to be isolated from the untrusted OS and Direct Memory Access~(DMA). 
Second, the trusted path itself requires isolation from the OS. 
Third, the user device needs isolation from all other devices which are under control of the OS. 
This section outlines how SGXIO meets these isolation requirements.

\textbf{Trusted Memory Isolation} is a prerequisite for securely executing trusted code in an untrusted environment. 
This affects user apps as well as the trusted stack. 
To achieve memory isolation of the user app, it is executed within an enclave. 
SGX isolates all enclave memory from the untrusted OS. 
To achieve memory isolation for the trusted stack, the hypervisor confines the untrusted OS in a VM. 
Moreover, the hypervisor implements a strict memory partitioning by configuring the Memory Management Unit~(MMU) appropriately. 
This prevents the OS from escaping the VM and tampering with the trusted stack. 

Direct Memory Access~(DMA) is a more subtle threat to memory isolation~\cite{zhou_building_2012}. 
A DMA-capable device can directly access memory, bypassing any MMU protection and potentially violating integrity and confidentiality of trusted memory. 
SGX prevents DMA from accessing enclave memory, hence the user app is safe against DMA attacks~\cite{sgxmanual}. 
Likewise, the trusted stack has to be protected against such attacks. 
Modern chipsets typically incorporate an I/O Memory Management Unit~(IOMMU), also termed \VTd{} on Intel systems. 
The IOMMU restricts device DMA to specific portions of RAM only. 
By properly configuring the IOMMU, the hypervisor can protect the whole trusted stack against device DMA attacks.

\textbf{Trusted Path Isolation.}
The trusted path has to be protected on two layers, namely the communication between user app and driver as well as the interaction between driver and user device. 
The user app communicates with the driver via the untrusted OS stack, hence encryption is necessary. 
The interaction between driver and user device is protected by the hypervisor. 
Therefore, the hypervisor establishes an exclusive binding between a driver and the corresponding user device. 
Moreover, the hypervisor mutually isolates all drivers. 
Thus, an attacker, loading arbitrary driver code at will, cannot interfere with trusted paths established by other drivers.

\textbf{User Device Isolation.}
As outlined before, a malicious OS could misconfigure devices to interfere with the trusted path. 
In that way, OS-controlled PCI devices could be forced to overlap their MMIO region or I/O port range with those of the user device or issue forged interrupts on behalf of the user device. 
To protect against these attacks, Zhou \etal implement several policies in the hypervisor to detect and prevent malicious device configurations. 
This effectively isolates a user device from other OS-controlled devices. 
Their approach is also applicable to SGXIO.

\subsection{User App Design}
Secure user applications play a central role in concrete use case scenarios like secure online banking. 
This section outlines principles for designing user apps and shows how user apps securely communicate with drivers. 
In the end, we elaborate on the enclave programming model. 

\textbf{Design Principles.}
Key to any secure user-centric application is a trusted path which protects user I/O against advanced malware like keyloggers and likewise. 
Without trusted path, an attacker could impersonate the user and act on its behalf, even if the user app itself is not compromised. 
To provide its service, a user app might communicate with other user apps or exchange sensitive data with a remote server using TLS, for example. 
Any operation on sensitive data is carried out within an enclave. 
Unproblematic code is kept outside the enclave. 
This covers glue code to the OS and untrusted libraries like file management, network socket access and likewise. 
All interaction with untrusted code needs careful validation inside the enclave~\cite{ports_towards_2008,checkoway_iago_2013,haven}. 
To keep state among multiple invocations, the user app can encrypt sensitive data for offline storage using SGX sealing, for example. 

\textbf{Encrypted Channel.}
To open a trusted path, the user app sets up an encrypted channel to a secure I/O driver. 
An encrypted channel protects sensitive user I/O against the untrusted OS. 
To open such a channel, the user app needs to share an encryption key with the driver via some form of key exchange. 
SGX local attestation can assist in key exchange by providing means to authentically exchange information between user app enclaves and driver enclaves. 
A straight-forward implementation uses Diffie-Hellman key exchange, as suggested by~\cite{anati_innovative_2013,sgxsdk}. 
However, local attestation inherently provides a much faster way of exchanging key material. 
We give a novel, lightweight key transport scheme, which comes with just a single uni-directional invocation of local attestation. 
We reuse the already pre-shared report key to derive random 128-bit encryption keys. 
The scheme works as follows: 
The user enclave generates a local attestation report over a random salt, targeted at the driver enclave. 
However, instead of delivering the actual report to the driver enclave, the user enclave keeps it private and uses the report's MAC as symmetric key. 
It then sends the salt and its identity to the driver enclave, which can recompute the MAC to obtain the same key. 
Details of this scheme are given in the Appendix. 

Once a key is established, one can use any authenticated encryption to ensure confidentiality and integrity of the data stream between user app and driver. 
Use of an authenticated encryption scheme ensures confidentiality and integrity of the data stream. 
By doing key exchange via SGX local attestation, the user app and the driver can mutually authenticate each other. 
This is also referred to as origin integrity. 

\textbf{Enclave Programming Model.}
SGXIO benefits from SGX's easy enclave programming model~\cite{sgxdevguide,sgxtutorial}. 
User app enclaves are executed directly on a host application running within the untrusted OS. 
Hence, secure user applications are treated similar to ordinary application processes. 
The OS has control over memory management, process management and scheduling of enclaves, although SGX carefully validates any action that might affect enclave security. 
Also, integration of multiple enclaves into a bigger user application stack is easy since enclaves share parts of the register set and the virtual address space with their host for communication purposes. 
Moreover, SGX supports multithreading as well as enclave debugging. 
Also, SGX natively provides verified launch and attestation of enclaves, which is tedious to implement in software. 
To support enclave development, Intel provides a software development kit~\cite{sgxsdk} as well as a comprehensive enclave developer guide~\cite{sgxdevguide}. 

\subsection{Driver Design}

Secure I/O drivers are responsible for connecting user apps and user devices.
Drivers are hosted and protected by the hypervisor. 
Although hypervisor protection is sufficient to isolate drivers from the untrusted OS, actual driver logic is in addition executed in an enclave. 
This helps in setting up an encrypted communication channel with user apps, as previously described. 
Also, driver enclaves are subject to attestation, allowing identification via their \verb!MRENCLAVE! values.

When designing a driver one has to make certain design choices. 
We opt for two strategies, namely domain multiplexing and portability, targeting commodity operating systems. 
Note that SGXIO supports other choices as well. 
\emph{Domain multiplexing} allows the same driver and thus the same user device to be shared across security domains. 
\emph{Portability} refers to drivers being compatible to different operating systems. 

\textbf{Domain Multiplexing.}
A driver handles the data stream from and to a user device and forwards it to the OS or a user app, respectively. 
Since many user devices like human interface devices or graphic cards are potentially shared between the untrusted OS and user apps, the driver has to multiplex the data stream between those security domains. 
In our example (see Figure~\ref{fig:sw_stack}), the driver offers its service to the OS via two separate virtual devices. 
During normal operation, the driver simply routes the unmodified data stream to the first virtual device, which matches the device class of the user device. 
This gives the OS transparent access to the user device. 
If the user app requests a trusted path, the driver redirects all traffic to the second virtual device, however in an encrypted fashion. 
The user app, knowing the proper decryption key, can access this second virtual device to tunnel through the untrusted OS. 
This second virtual device can be any standard character device, for example, which just forwards the encrypted data stream. 
In this example, the driver implements strict temporal multiplexing between the OS and the user app. 
However, one could enforce arbitrary security policies. 
For example, the driver could implement spatial partitioning of a graphic card's frame buffer to allow secure screen overlays. 
Or it could intercept and mask certain keystrokes to react on secure attention sequences~\cite{perrig_safe_2009} and encrypt password entry, for example.

\textbf{Portability.}
In our example we encourage virtual devices as communication interface between drivers and the OS, \cf Figure~\ref{fig:sw_stack}. 
This has the advantage of being completely compatible to commodity OSes. 
No changes to the OS are required since a user device is perfectly emulated by the driver. 
Also, since the driver has no notion of which OS it is serving, one and the same driver implementation can be reused across multiple different OSes without porting effort. 
Saved manpower can be put in a robust driver implementations. 
Note that specific high-throughput user devices might need cooperation by the OS, breaking full portability. 

\subsection{Hypervisor Design}

The hypervisor is responsible for running the untrusted OS in a VM as well as loading drivers and binding user devices to them. 
Drivers can be statically loaded by the hypervisor on system boot. 
This makes sense for permanently installed user devices like notebook keyboards and graphic cards. 
Drivers for plug-and-play devices like USB might be dynamically loaded by the hypervisor. 
Note that typically the hypervisor delegates such resource manangement tasks to a separate Virtual Machine Monitor~(VMM). 

The hypervisor enforces a bunch of isolation guarantees, as previously outlined: 
First, it isolates all trusted stack memory. 
Second, it binds a user device exclusively to the corresponding driver and mutually isolates drivers. 
This achieves trusted path isolation. 
Third, it isolates user devices from malicious interference with other devices. 

\textbf{seL4.}
Choice of an appropriate hypervisor fulfilling these requirements is crucial for overall system's security.
We recommend to use seL4 as hypervisor, as it allows a straight forward design of SGXIO. 
seL4 implements a strict resource partitioning, which directly supports isolation of trusted memory as well as user device binding. 
Therefore, seL4 knows capabilities for each resource~\cite{sel4manual2.0}. 
By granting the VM or a driver specific capabilities, it gets access to the underlying resources. 
With such a capability system in place, isolation breaks down to a correct distribution of memory and device capabilities among the VM and the drivers. 
For example, the VM as well as each driver gets assigned a disjoint set of memory capabilities, enforcing memory isolation. 
Likewise, each driver gets capabilities to its own user device only. 
Capabilities to other devices are given to the VM. 
This enforces trusted path isolation. 

To enforce its capability system, seL4 makes heavy use of Intel's \VTx{} hardware virtualization. 
Each illegitimate memory or device access, be it via ordinary memory addressing, Memory-Mapped I/O~(MMIO) or I/O ports, is intercepted by means of \VTx{}. 
In the same way, \VTx{} helps in blocking device misconfiguration attacks~\cite{zhou_building_2012} and achieving user device isolation. 
Furthermore, seL4 uses Intel \VTd{}, also referred to as IOMMU,  to protect against DMA attacks from misconfigured devices. 
Thus, seL4 is perfectly suitable to implement all of the above isolation guarantees.

Moreover, seL4 is formally verified~\cite{sel4proof,murray_sel4:_2013}. 
The proofs not only cover functional correctness of the generic C implementation but also help finding a correct kernel configuration under which isolation guarantees hold. 
The developers of seL4 claim to have the first general-purpose microkernel with such strong guarantees. 
Although initial proofs were conducted for the ARM architecture, most verified, generic code is shared with x86.  

As with seL4, XMHF aims at strong memory isolation, backed by formal proofs~\cite{vasudevan_design_2013}. 
Using a formally verified hypervisor like seL4 or XMHF has several advantages. 
The hypervisor is essential part of the TCB, responsible for establishing the generic trusted path. 
The formal proofs help making strong claims about security of the trusted path, which is not the case when relying on a best-effort implementation without such proofs. 
Once the hypervisor is considered trustworthy, subsequent reasoning can concentrate on single code modules like drivers rather than the whole system. 


\section{Domain Binding}\label{sec:security}

This section elaborates on challenges which arise when binding the SGX domain with the trusted hypervisor domain. 
Specifically, this covers trusted boot and hypervisor attestation. 
We discuss how to protect hypervisor attestation against remote TPM attacks as well as enclave virtualization attacks. 
Having a domain binding in place allows remote attestation of trusted paths as well as user verification.

\subsection{Challenges}

SGXIO enables a remote party as well as a local user to verify security of trusted paths. 
In the first place, this requires a domain binding between SGX and the trusted hypervisor. 
In the second place, an appropriate user verification mechanism needs to be in place which is both secure and usable.

\textbf{Domain Binding.}
In order to bind the SGX and the hypervisor domain, the hypervisor must level up to certain security guarantees SGX regarding isolated code execution. 
In SGX, all enclave memory is isolated from the rest. 
Moreover enclave loading is guarded by a verified launch mechanism, which can be attested to other parties. 
SGXIO  rebuilds similar mechanisms for the hypervisor. 
Isolation of trusted memory has already been discussed in Section~\ref{sec:isolation}. 
Verified launch is implemented via \emph{trusted boot} of the hypervisor with support for \emph{hypervisor attestation}. 

With trusted boot and hypervisor attestation in place, SGXIO can bind the SGX and the hypervisor domain. 
The binding needs to be bidirectional, allowing both the hypervisor and SGX enclaves to put trust in the other domain. 
One direction is easy: The hypervisor can extend trust to SGX by running enclaves in a safe, hypervisor-protected environment. 
These enclaves can in turn use local attestation to extend trust to any other enclaves in the system. 
However, the opposite direction is challenging: 
On the one hand, enclaves need confidence that the hypervisor is not compromised and binds user devices correctly to drivers. 
Effectively, this requires enclaves to invoke hypervisor attestation. 
SGXIO achieves this with assistance of the TB enclave.
On the other hand, SGX is not designed to cooperate with a trusted hypervisor. 
Recall that SGX considers all non-enclave code untrusted. 
In fact, SGX explicitly prohibits use of any instruction inside an enclave that might be used to communicate with the hypervisor~\cite{sgxdevguide}. 
Even if hypervisor attestation succeeds, an enclave cannot easily learn whether it is legitimately executed by the hypervisor or virtualized by an attacker in a fake environment. 
This makes driver enclaves and the TB enclave vulnerable to virtualization attacks. 
SGXIO defends against such attacks by hiding hypervisor attestation from the untrusted OS.

\textbf{User Verification.}
An end user wants to be able to verify if he is indeed communicating with the correct user app via a trusted path. 
This is non-trivial because the user cannot simply evaluate an cryptographic attestation report. 
Instead, the user requires some form of notification whether a trusted path is present. 
This notification needs to be unforgeable to prevent the OS from faking it. 
Moreover, it needs to help the user distinguish different user apps, not least because an attacker might also run arbitrary user apps under his control.

\subsection{Trusted Boot \& Hypervisor Attestation}\label{sec:hypervisor_attest}
Trusted boot allows verifying integrity of the hypervisor by doing a measured launch over all booted code. 
Without it, malware could silently hook into the boot process and disable any protection offered by the hypervisor. 
In contrast to verified launch, measured launch does not prohibit booting a compromised hypervisor but records it for later evaluation. 

Trusted boot makes use of a TPM to measure the whole boot process, starting from a trusted piece of firmware code up to the hypervisor image. 
During a normal boot, each boot stage measures the next one in a cryptographic log inside the TPM using the \texttt{extend} operation. 
All measurements are cumulated in a TPM Platform Configuration Register~(PCR). 
The final PCR value reflects the whole boot process. 
If any boot stage deviates from the normal boot process, the PCR will contain a wrong value.

\textbf{Hypervisor Attestation}
allows enclaves to verify the trusted boot process in order to get assured of hypervisor's integrity. 
Since the hypervisor is responsible for loading drivers and doing trusted path setup, its attestation also vouches for security of all trusted paths. 

\begin{figure*}[t]
\centering
\includegraphics[width=1\textwidth]{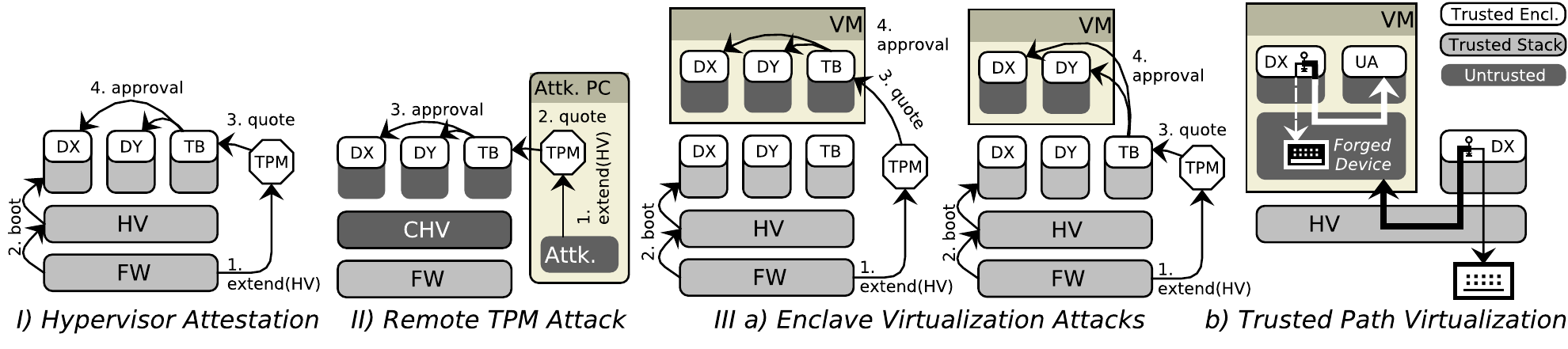}
\caption{I) During trusted boot, Firmware~(FW) measures the Hypervisor~(HV) via a TPM. 
The TB enclave attests the hypervisor via a TPM \texttt{quote} and in case of success, approves other drivers (DX) and (DY). 
II) An attacker injects a remotely-generated TPM \texttt{quote} to hide the presence of a Compromised Hypervisor~(CHV). 
IIIa) An attacker diverts steps 3 or 4 to a virtualized environment. 
IIIb) This allows to virtualize a trusted path to a user app~(UA).}
\label{fig:hv_attack}
\end{figure*}

To ease hypervisor attestation, SGXIO uses a Trusted Boot~(TB) enclave which attests the hypervisor once. 
Afterwards, any driver enclave running on the system can query the TB enclave to get approval if hypervisor attestation succeeded, see Figure~\ref{fig:hv_attack}/I. 
The driver enclaves in turn can communicate the attestation result to user apps, which can finally implement a mechanism for remote parties or the end user to verify a trusted path.

To attest the hypervisor, the TB enclave needs to verify the PCR value, obtained during trusted boot. 
Therefore, the TB enclave requests a TPM \texttt{quote}~\cite{tpm}, which contains a cryptographic signature over the PCR value alongside with a fresh nonce. 
This ensures not only integrity of the PCR value but also prevents replay attacks. 
 
\subsection{Attacks}
 The interaction between TB enclave and TPM is crucial for security of the hypervisor attestation scheme. 
One has to prevent remote TPM attacks as well as enclave virtualization attacks, which is outlined in the following.
 
\subsubsection{Remote TPM attacks.}

If the TB enclave does not identify the TPM it is talking to, hypervisor attestation becomes vulnerable to remote TPM attacks, also called cuckoo attacks~\cite{parno_bootstrapping_2008}. 
If the attacker compromises the hypervisor image, the PCR will yield a wrong value during trusted boot, which is detected by the TB enclave. 
However, the attacker can make hypervisor attestation work again by diverting TB enclave communication to an attacker-controlled TPM. 
Since the attacker can feed the remote TPM at will to generate a valid \texttt{quote}, the TB enclave successfully approves the compromised hypervisor. 
See also Figure~\ref{fig:hv_attack}/II. 

\textbf{Defense.}
In order to stop cuckoo attacks, the TB enclave needs to be bound to a particular computer. 
In other words, the TB enclave needs a-priori knowledge of the TPM, e.g., in form of the TPM's Attestation Identity Key~(AIK) used for signing the \texttt{quote}. 
This allows the TB enclave to verify the origin of the TPM \texttt{quote}. 
To make the AIK known to the TB enclave, one has to provision it to the TB enclave, e.g., during initial system integration. 
The TB enclave stores the provisioned AIK together with some redundancy by means of SGX sealing. 
On hypervisor attestation, the TB enclave unseals the AIK and uses it to verify the \texttt{quote}. 
Redundancy helps in verifying integrity of the unsealed AIK.
Since sealing uses a CPU-specific encryption key, an attacker cannot trick the TB enclave to unseal an AIK not sealed by the same CPU. 
This effectively binds execution of the TB enclave to the TPM.

Provisioning of AIKs could be done by system integrators. 
One has to introduce proper measures to prevent attackers from provisioning arbitrary AIKs. 
For example, the TB enclave could encode a list of public keys of approved system integrators, which are allowed to provision AIKs.

\subsubsection{Enclave Virtualization Attacks}\label{sec:driver_virtualization}
SGX is not designed to cooperate with a trusted hypervisor, making driver enclaves as well as the TB enclave vulnerable to enclave virtualization attacks. 
In an enclave virtualization attack, the attacker does not compromise the actual trusted boot process. 
Rather, he virtualizes driver enclaves or even the TB enclave in a fake environment on the same computer, as depicted in Figure~\ref{fig:hv_attack}/IIIa. 
To make hypervisor attestation for the virtualized enclaves succeed, the attacker diverts the legitimate TPM quote or the TB enclave approval to the virtualized TB enclave or driver enclaves, respectively. 
As shown in Figure~\ref{fig:hv_attack}/IIIb, the attacker can now impersonate the user by rerouting user apps to a virtualized driver, reading driver's output and providing fake input. 
Note that the attacker did not change enclave code. 
Hence, SGX will generate the same \verb!MRENCLAVE! value and thus the same derived cryptographic keys for both, legitimate and virtualized enclave instances. 
Neither the TB enclave nor the driver enclave or a user app can detect such virtualization. 
However, the attacker does not learn actual user input, which still arrives at the legitimate driver enclave. 

\textbf{Defense.}
The problem of enclave virtualization stems from the design of SGX which treats all enclaves equally, regardless of the security context they are executed in. 
As a defense, SGXIO restricts the communication interface between the hypervisor and the OS context. 
Therefore, the hypervisor hides the TPM as well as the TB enclave from the untrusted OS\footnote{This breaks path 3 on the left and path 4 on the right side of Figure~\ref{fig:hv_attack}/IIIa, respectively.}. 
Only the legitimate TB enclave is given access to the TPM. 
Thus, the TB enclave might only succeed in hypervisor attestation if it has been legitimately launched by the hypervisor. 
Likewise, only legitimate driver enclaves are granted access to the legitimate TB enclave by the hypervisor. 
A driver enclave might only get approval if it can talk to the legitimate TB enclave, which implies that the driver enclave too has been legitimately launched by the hypervisor. 

Note that user app enclaves are not subject to enclave virtualization attacks since they are already running in the OS context and do notexchange sensitive plain data with their untrusted environment.

\subsection{Remote Trusted Path Attestation}

\begin{figure}
\centering
\includegraphics[width=0.45\textwidth]{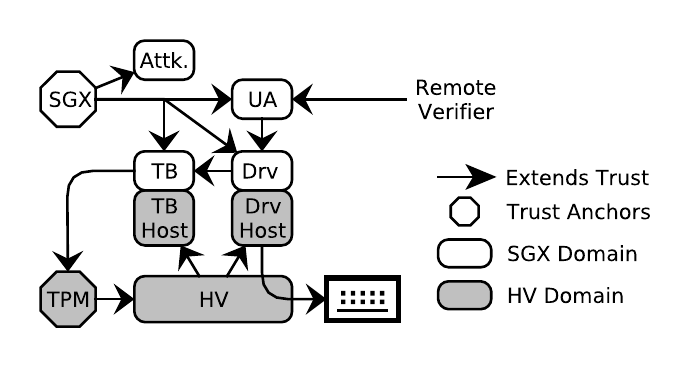}
\caption{Trust hierarchy of SGXIO.}
\label{fig:trust_chain}
\end{figure}

As already mentioned, hypervisor attestation vouches for security of trusted paths and serves as basis for remote attestation. 
This section describes the whole trust hierarchy involved in remote attestation, as shown in Figure~\ref{fig:trust_chain}.

SGXIO has two main hardware trust anchors, namely SGX and the TPM. 
SGX extends trust to all enclaves running on the system by means of verified launch. 
This also includes enclaves with attacker-controlled code~(Attk.). 
It is up to a remote verifier and individual enclaves to build a trust hierarchy among "good" enclaves. 
To do this, trust is extended via SGX remote and local attestation, respectively. 
The entrusting party does not only verify validity of an SGX attestation report but also check for a correct \verb!MRENCLAVE! value, which uniquely identifies an enclave codebase. 

In a typical scenario, a remote verifier wants to establish a trusted path to a user. 
It therefore extends trust to a specific user app~(UA) under its control, which in turn entrusts appropriate secure I/O drivers~(Drv). 
Drivers extend trust to the TB enclave, which does hypervisor attestation as previously outlined. 
If hypervisor attestation succeeds, trust is implicitly extended to the TB host and all driver host processes together with all trusted paths to user devices. 
If at any point in the trust hierarchy attestation fails, the affected entities will terminate trusted path attestation. 

\subsection{User Verification}\label{sec:user_verify}
SGXIO allows a user to locally assess if he is talking with the correct user app via a trusted path. 
This does not require additional hardware such as an external handheld verification device~\cite{zhou_building_2012} or similar. 
Instead, we stick to sharing a secret piece of information between user and user app, similar to~\cite{verifiedbyvisa}. 
For the sake of simplicity we discuss the common scenario of a trusted screen path and a trusted keyboard path. 
When the user starts the user app, the user app requests a trusted input path to the keyboard and a trusted output path to the screen from the corresponding drivers. 
If for any reason one or both trusted path setups fail, the user app terminates with an error. 
In the case of success, the user app displays the pre-shared secret information via the screen driver to the user. 
The user verifies this information to get assured of a valid trusted path setup for this user app. 
Since an attacker does not know the secret information, he cannot fake this notification.
This approach requires provisioning secret information to a user app, which seals it for later usage. 
Provisioning could be done once at installation time in a safe environment, \eg with assistance of the hypervisor, or at any time via SGX's remote attestation feature. 


\section{Tweaking Debug Enclaves}\label{sec:sgx_debug}
Our architecture makes heavy use of SGX enclaves. 
In order to enable full security of production enclaves, SGX enforces a licensing scheme on enclave code, which might be too costly for small business or even incompatible with the open-source idea. 
In this section we show how to level up debug enclaves to behave like production enclaves in the threat model of SGXIO. 
This requires special handling of SGX remote attestation and sealing. 
Note that the debug tweak does not apply to a secure cloud computing scenario with an untrusted cloud provider. 

Recall that the only difference between debug and production enclaves is the presence of SGX debug instructions, which we aim to disable manually.
The debug tweak leverages SGX's support for \VTx{} instruction interception. 
\VTx{} supports several configurable bitmaps, making certain instructions trap into the hypervisor when executed from withing a VM. 
These bitmaps are configured by the hypervisor via so-called Virtual Machine Control Structures~(VMCS). 
With the release of SGX, Intel added a new bitmap for SGX ENCLS instructions, called ENCLS-exiting bitmap. 
This allows the hypervisor to selectively intercept ENCLS instructions. 
Thus, the hypervisor can intercept all \verb!EDBGRD! and \verb!EDBGWR! instructions which are ever executed from within a VM, by just configuring the VMCS bitmaps accordingly. 
By doing so, the only code which is able to debug enclaves is the trusted hypervisor itself. 
Since the hypervisor is trusted, we can consider SGX debugging features as disabled. 
Hence, we have effectively turned all debug enclaves inside the VM into production equivalents.

\textbf{Tweaked Cloud Enclaves.}
As already mentioned, this tweak only applies to a setting similar to SGXIO, where a trusted hypervisor is present. 
In general, this is not the case for cloud scenarios where the cloud provider is untrusted and expected to subvert the hypervisor. 
In such cases, one has to opt for real production enclaves. 
Nevertheless, honest server administrators could use the tweak to obtain SGX protection without licensing. 
This would help in strongly isolating server code and reducing the TCB from the whole system down to the hypervisor and the enclave code.

\textbf{Remote Attestation.}
With tweaked debug enclaves, remote attestation requires special care since a remote verifier cannot easily determine whether the debug tweak is correctly enabled or not. 
For example, an attacker could compromise the hypervisor and manipulate (debug) the TB enclave to issue wrong approvals. 
Next, the attacker could stealthily debug all enclaves on the system. 

To do remote attestation with tweaked debug enclaves, one can run only the TB enclave in production mode and do remote attestation towards it. 
Once a remote party verified the TB enclave, it can be sure that the hypervisor correctly enforces the tweak for all debug enclaves in the system.

\textbf{Sealing.}
Both, non-tweaked and tweaked debug enclaves share the same sealing keys. 
This is no problem unless an attacker manages to compromise the hypervisor and disables the tweak. 
Although hypervisor attestation would fail in that case, the attacker would be able to extract all sealing keys by simply debugging all enclaves. 
To prevent this, one can delegate sealing key derivation to the TB enclave. 
The TB enclave, running in production mode, only derives actual sealing keys if hypervisor attestation succeeds.


\section{Further Considerations}\label{sec:implementation}
This section touches on advanced topics potential users of SGXIO should be aware of, namely driver complexity and side-channels. 

\textbf{Driver Complexity.} 
Depending on the bus protocol, driver design might be challenging. 
Especially multiplexed buses, such as USB are non-trivial to deal with. 
One has to identify proper policy rules which guarantee a trusted path. 
Zhou \etal demonstrate how to establish a trusted path to one specific USB device, while keeping other USB devices accessible to the untrusted OS~\cite{zhou_dancing_2014}. 
To deal with complexity of the USB driver stack, they identified all security-relevant parts and either moved them entirely into a trusted domain or at least verified their results. 
All non-critical operations are kept in the untrusted domain. 
This approach would in principle also be supported by SGXIO with a cooperative design of the OS and secure I/O drivers.

\textbf{SGX Side-Channels.}\label{sec:paging_sidechannel}
The threat models of SGX and SGXIO do not consider side channels~\cite{sgxdevguide}. 
SGX is vulnerable to a paging side channel, leaking address information on a page granularity~\cite{controlledchannel}. 
Likewise, SGX is most likely susceptible to timing attacks on the cache or DRAM~\cite{costan_intel_2016,pessl_drama:_2016}. 
To mitigate such side channels, enclave developers have to stick to implementation-level approaches. 
This includes reduction of sensitive data dependencies and hiding of memory access patterns~\cite{controlledchannel,ohrimenko_oblivious_2016,sgxtutorial}. 


\section{Conclusion}\label{sec:conclusion}

We present SGXIO, the first SGX-based architecture to support generic trusted paths. 
Therefore, we augment SGX with a small and trusted hypervisor for setting up a generic trusted path, while SGX helps in protecting user apps from an untrusted OS. 
We solve the challenge of combining the security domains of SGX and the hypervisor. 
We do so by attesting the hypervisor with assistance of a TPM towards a special trusted boot enclave, which is bound to the local computer. 
With SGXIO, both a remote party and a local user can verify security of trusted paths.

SGXIO has an easy programming model, allowing application developers to integrate sensitive code directly in existing, untrusted code and open a trusted path by means of ordinary, virtual device I/O. 
Furthermore, we show how SGXIO can omit enclave licensing by making debug enclaves behave like production enclaves. 
Therefore, the trusted hypervisor disables SGX debugging instructions for the whole untrusted VM. 

SGXIO demonstrates that SGX is not limited to cloud computing and DRM scenarios. 
SGXIO addresses user-centric application security, making generic trusted paths available to SGX enclaves. 
This can greatly improve application security, protecting against kernel-level keyloggers and screenloggers, for example. 
Moreover, SGXIO can help honest web administrators to isolate sensitive web services within a tweaked debug enclaves without the need for enclave licensing. 
SGXIO is compatible to unmodified legacy OSes and runs on off-the-shelf notebooks.


\section*{Acknowledgments}
This work was partially supported by the TU Graz LEAD project "Dependable Internet of Things in Adverse Environments".




\newif\ifabfull\abfullfalse
\input apreambl \newif\ifger\gerfalse \ifnum\language=\german\gertrue\fi
  \ifnum\language=\austrian\gertrue\fi \def\sortby#1{}


\appendix\label{sec:appendix}

\balance

We present a fast non-interactive key transport scheme for local SGX enclaves based on local attestation. 
First, local attestation is outlined. 
Second, our scheme is given in detail.

Local attestation uses the two instructions \verb!EREPORT! and \verb!EGETKEY! to generate and verify an attestation report, respectively. 
Both instructions are outlined in Algorithm~\ref{alg:ereport} and~\ref{alg:egetkey}. 
To attest itself to a target enclave T, enclave E generates a report using \verb!EREPORT!. 
To verify the report, enclave T queries the report key via \verb!EGETKEY! and manually re-calculates the report signature. 
The local attestation procedure is outlined in Algorithm~\ref{alg:local_attestation}. 
For the sake of simplicity, we omit some details in the algorithm descriptions. 
Furthermore, we denote \verb!MRENCLAVE! of enclave X as $X_{ID}$, as it serves as enclave identifier.

\balance

\vspace{0.8em}
\begin{algorithm}[H]
\floatname{algorithm}{Algorithm}
\begin{algorithmic}
\STATE \verb!EREPORT! generates a report of enclave E, targeted at enclave $T$, containing additional user $data \in \{0,1\}^{256}$.
\STATE $rpkey_{T_{ID}} \leftarrow SHA_{256}(T_{ID})$
\STATE $mac \leftarrow AES_{CMAC}(rpkey_{T_{ID}}, E_{ID} || data)$
\STATE return $(E_{ID} || data || mac)$
\end{algorithmic}
\caption{$EREPORT_{E_{ID}}(T_{ID}, data)$}
\label{alg:ereport}
\end{algorithm}
\vspace{0.8em}

\begin{algorithm}[H]
\floatname{algorithm}{Algorithm}
\begin{algorithmic}
\STATE \verb!EGETKEY! returns the report key of enclave E.
\IF {$key\_type = REPORT\_KEY$}
\STATE return $SHA_{256}(E_{ID})$
\ENDIF
\end{algorithmic}
\caption{$EGETKEY_{E_{ID}}(key\_type)$}
\label{alg:egetkey}
\end{algorithm}
\vspace{0.8em}

\begin{algorithm}[H]
\floatname{algorithm}{Algorithm}
\begin{algorithmic}
\STATE Enclave $E$ attests itself to target enclave $T$, providing additional $data$ to be authenticated.
\STATE $E\rightarrow{T}$ : $rp \leftarrow EREPORT_{E_{ID}}(T_{ID}, data)$
\STATE $T$ : $rpkey_{T_{ID}} \leftarrow EGETKEY_{T_{ID}}(REPORT\_KEY)$
\STATE $T$ : $mac \leftarrow AES_{CMAC}(rpkey_{T_{ID}}, rp.E_{ID} || rp.data)$
\IF {$mac = rp.mac$} \STATE $T$ : accept $rp$ \ELSE \STATE $T$ : reject $rp$ \ENDIF
\end{algorithmic}
\caption{$LocalAttestation_{E_{ID}}(T_{ID}, data)$}
\label{alg:local_attestation}
\end{algorithm}

\textbf{Non-Interactive Key Transport Scheme.}
SGX uses an AES-based CMAC~\cite{aes_cmac}. 
This is a symmetric signature scheme using the same key for signature creation and verification. 
Hence, local attestation already provides a shared symmetric secret between enclaves, namely the report key. 
Any enclave can sign reports for a target by issuing \verb!EREPORT!. 
Although it has no direct access to the target enclave's report key, it can indirectly use it via the \verb!EREPORT! instruction. 
In turn, the target enclave is able to access its own report key via \verb!EGETKEY!. 
We use this symmetry of report keys to derive fresh encryption keys. 

To establish a new key between enclave A and B, A chooses a random nonce and generates an attestation report over this nonce, targeted at B. 
However, A never transmits this report to B but uses the report's MAC as an 128-bit shared, symmetric encryption key. 
Instead, enclave A sends its identity as well as the nonce to B, which can query its report key and re-calculate the MAC to obtain the shared key. 
The scheme is outlined in Algorithm~\ref{alg:symmetric_keyexchange}. 

\begin{algorithm}[H]
\floatname{algorithm}{Algorithm}
\caption{Non-interactive, symmetric key exchange}
\begin{algorithmic}
\STATE Enclave $A$ sends a fresh symmetric key to enclave $B$.
\STATE $A$ : $nonce \overset{R}{\leftarrow} \{0,1\}^{256}$ chosen uniformly at random
\STATE $A$ : $rp \leftarrow EREPORT_{A_{ID}}(B_{ID}, nonce)$
\STATE $A$ : $key_{AB} \leftarrow rp.mac$
\STATE $A\rightarrow{B}$ : $(A_{ID}, nonce)$
\STATE $B$ : $rpkey_{B_{ID}} \leftarrow EGETKEY_{B_{ID}}(REPORT\_KEY)$
\STATE $B$ : $key_{AB} \leftarrow AES_{CMAC}(rpkey_{B_{ID}}, A_{ID} || nonce)$
\end{algorithmic}
\label{alg:symmetric_keyexchange}
\end{algorithm}

Our scheme is non-interactive since it only involves a single uni-directional transmission of the nonce. 
It has zero overhead since local attestation is supported by SGX hardware. 
The only noteworthy enclave code, namely the AES-CMAC implementation, is typically already part of an enclave codebase for doing local attestation. 
Moreover, the scheme supports authentication. 
On the one hand, enclave A binds the key to enclave B by means of local attestation. 
On the other hand, enclave B knows the identity of the enclave it is deriving a shared key for. 
To also achieve liveness, enclave B could send the encrypted nonce back to A.

\balancecolumns 
\end{document}